# Surface Diffusion Control Enables Tailored Aspect Ratio Nanostructures in Area-Selective Atomic Layer Deposition

Philip Klement[1], Daniel Anders[1], Lukas Gümbel[1], Michele Bastianello[1], Fabian Michel[1], Jörg Schörmann[1], Matthias T. Elm[1,2], Christian Heiliger[3], and Sangam Chatterjee[1,*]

[1] Institute of Experimental Physics I and Center for Materials Research (ZfM/LaMa), Justus Liebig University Giessen, Heinrich-Buff-Ring 16, Giessen, D-35392, Germany

[2] Institute of Physical Chemistry, Justus Liebig University Giessen, Heinrich-Buff-Ring 17, Giessen, D-35392, Germany

[3] Institute of Theoretical Physics and Center for Materials Research (ZfM/LaMa), Justus Liebig University Giessen, Heinrich-Buff-Ring 16, Giessen, D-35392, Germany

Area-selective atomic layer deposition is a key technology for modern microelectronics as it eliminates alignment errors inherent to conventional approaches by enabling material deposition only in specific areas. Typically, the selectivity originates from surface modifications of the substrate that allow or block precursor adsorption. The control of the deposition process currently remains a major challenge as the selectivity of the no-growth areas is lost quickly. Here, we show that surface modifications of the substrate strongly manipulate the surface diffusion. The selective deposition of $TiO_2$ on poly (methyl methacrylate) and $SiO_2$ yields localized nanostructures with tailored aspect ratios. Controlling the surface diffusion allows to tune such nanostructures as it boosts the growth rate at the interface of the growth and no-growth areas. Kinetic Monte-Carlo calculations reveal that species move from high to low diffusion areas. Further, we identify the catalytic activity of $TiCl_4$ during the formation of carboxylic acid on poly (methyl methacrylate) as the reaction mechanism responsible for the loss of selectivity, and show that process optimization leads to higher selectivity. Our work enables the precise control of area-selective atomic layer deposition on the nanoscale, and offers new strategies in area-selective deposition processes by exploiting surface diffusion effects.

INTRODUCTION

The downscaling of integrated circuits in semiconductor technology currently relies almost completely on top-down processing featuring a complex combination of many material deposition, extreme ultraviolet lithography and etching steps. Approaching sub-5 nm device scales makes perfect alignment in nanopatterning for reliable processing extremely challenging as it mandates virtually atomic-scale precision.[1] In particular, alignment-related edge placement errors have become a major issue in downscaling.[2] Consequently, new bottom-up schemes for the most-demanding processing steps are required to circumvent challenges or even avoid inherent deficiencies of top-down fabrication.

Atomic layer deposition (ALD) is such a bottom-up thin film deposition technique where atoms or small units grow on a substrate layer-by-layer. It relies on exposing the substrate to alternating cyclic exposure of two or more vapor phase precursors that react in self-limiting surface reactions between functional groups and vapor phase precursors. This leads to growth with atomic-level thickness control and high conformity.[3] However, the bottom-up, self-aligned fabrication by ALD mandates the deposition processes limited to specific areas. As a result, area-selective atomic layer deposition (AS-ALD) emerged over the recent years.[4] This approach enacts spatial control of the growth areas by locally tailoring the specific surface chemistry. It enables bottom-up fabrication schemes that offer the potential to significantly reduce the number of required device processing steps. AS-ALD typically includes a substrate-patterning step before the area-selective ALD process. This reflects the reality of self-aligned fabrication schemes where pre-patterned and pre-processed materials serve as starting substrates. Such a patterning-centered reasoning classifies two main routes for AS-ALD: (A) AS-ALD by area passivation. Parts of the substrate surface are deactivated, *i.e.*, made inert to the chemistry of the ALD process; and (B) AS-ALD by area activation. Here, parts of the inert substrate surface are made reactive to the chemistry of the ALD process. Thus, a sample surface in AS-ALD features several materials and ALD deposition needs to be selective on one of these materials.

Previous work in the field of AS-ALD has covered dielectric-on-dielectric[5–7] and metal-on-dielectric[8–13] deposition as those are industrially most relevant. AS-ALD by area activation has been achieved by deposition of a metallic or oxidic seed layer,[5,9] whereas in AS-ALD by area deactivation, the substrate surface has been chemically modified by self-assembled monolayers (SAM),[7,8,13] and polymers.[6,10,12] Most AS-ALD studies focused on the demonstration of selectivity in a specific ALD process and the improvement of the selectivity towards thicker layers. Promising approaches such as the introduction of correction steps in so-called supercycles have been developed and enable high selectivity,[11,14] but the general problem of a loss of selectivity persists. Those studies have shown that the loss of selectivity involves nucleation on the no-growth area, but the nucleation process and limiting factors have not been clarified yet. They mandate the need for a better understanding of the fundamental interactions between precursors and reactants with specific surfaces in AS-ALD processes to enable higher selectivity.[2,4]

In this work, we successfully control the surface diffusion in the model system $TiO_2$ on poly (methyl methacrylate) (PMMA) to achieve selectively high aspect ratio nanostructures, and show that surface diffusion boosts the growth rate and leads to pile-up at the growth to no-growth interface. We use PMMA for area deactivation. It is a widely used and well-established polymer, which allows for nanometer-scale structures with ease of preparation. Further, it serves as a model system for AS-ALD processes based on blocking layers including SAMs and ABC-type supercycles with inhibitors as it features the same $CH_3$ groups that block precursor adsorption and thus are unreactive toward most ALD chemistries [11,14,15]. In the patterning step, a small lateral pattern establishes the basis for the AS-ALD process (Figure 1a). The use of electron beam lithography with its ability to generate sub-10 nm lateral patterns meets the resolution requirement of modern semiconductor device fabrication. We analyze the key factors in AS-ALD in terms of nucleation delay and selectivity measured by X-ray reflectivity (XRR), and find a selectivity window of 50 cycles or 3.12 nm with a unity selectivity for $TiO_2$ on PMMA. Further, we identify the catalytic activity of $TiCl_4$ during the formation of carboxylic acid on PMMA as the chemical reaction responsible for the loss of selectivity. We then elaborate on the selectivity on a molecular-scale level by atomic force microscopy (AFM), and find small islands of $TiO_2$ growing on PMMA beyond the selectivity window. On pre-patterned substrates, diffusion of species on the surface of PMMA results in a strongly increased growth per cycle (GPC) on $SiO_2$ compared to unpatterned substrates and a pile-up at the growth to no-growth interface. A kinetic Monte-Carlo model for particle diffusion excellently reproduces our experimental observations and identifies surface diffusion of species as the physical origin of both, the strongly increased GPC and the pile-up in AS-ALD. Finally, we control the surface diffusion of species through the purge times in the ALD process to enable the preparation of nanostructures with tailored aspect ratios and film thicknesses beyond the inherent selectivity of the AS-ALD process.

EXPERIMENTAL SECTION

**Substrate Preparation and Patterning Step.** Si substrates with a 275 nm thick wet thermal oxide layer were cleaned with acetone, isopropanol (IPA) as well as deionized (DI) water and coated with a 300 nm thick PMMA (950k molecular weight, 4 % dissolved in anisole) layer. Microscale patterns were defined by electron beam lithography (JEOL JSM 7001F electron microscope equipped with a XENOS XeDraw high speed writer system) followed by the development in a mixture of IPA and DI water (ratio 2:1). In some samples, $SiO_2$ was removed from the unexposed regions by reactive Ar-ion etching to form a trench.

**Atomic Layer Deposition and Lift-off.** TiO$_2$ was deposited in a commercial thermal ALD system (Picosun R200 Standard) at 120 °C to prevent the PMMA patterns from reflowing. Titanium tetrachloride and water were used as precursors and pulsed for 0.1 and 3 s, respectively (Figure S1). The system was purged with nitrogen for 10 s after each precursor pulse. After ALD, samples were immersed in 50 °C hot acetone for lift-off followed by cleaning with IPA and O$_2$ plasma for 3 minutes to remove residual polymer.

**Analytical methods.** The nucleation and selectivity were analyzed in terms of TiO$_2$ thicknesses, which were determined by X-ray reflectivity with a Siemens/Bruker D5000 X-ray diffraction system and a subsequent fit of the reflectivity (Figure S2). We use the generally accepted definition of the selectivity $S$ from the field of area-selective chemical vapor deposition

$$S = \frac{\theta_{GA} - \theta_{NGA}}{\theta_{GA} + \theta_{NGA}} \quad (1)$$

where $\Theta_{GA}$ and $\Theta_{NGA}$ are the amounts of material deposited (*i.e.*, in terms of film thickness, coverage, *etc.*) on the growth and no-growth areas, respectively.[16] Infrared (IR) absorption spectra were recorded on IR-transparent Si substrates with a Bruker Equinox 55 FTIR spectrometer equipped with a reflectance-measuring unit and converted to absorbance. We calculate the differential absorbance following

$$\Delta A_{n+25,n} = \frac{A_{n+25} - A_n}{A_n} \quad (2)$$

where $A_n$ is the absorbance of the sample with *n* cycles of TiO$_2$ (Figure S4). The topography of the TiO$_2$ patterns prepared by AS-ALD was analyzed with an AIST-NT SmartSPM 1000 AFM in noncontact tapping mode to prevent sample damage. We define the area ratio AR as

$$AR = \frac{A_{NGA}}{A_{GA}} \quad (3)$$

where $A_{GA}$ and $A_{NGA}$ are the absolute areas of the growth and no-growth areas, respectively.

**Kinetic Monte-Carlo Methods.** All molecular dynamics calculations were performed using a kinetic Monte-Carlo (KMC) approach. The model derives from a Markov-Chain approach and takes particle collision and occupied sites into account. Input parameters are lattice dimensions, iteration time, number of precursor molecules, size as well as proportion of different diffusion areas and the ratio of the different diffusion velocities.

RESULTS AND DISCUSSION

**Selectivity of $TiO_2$ ALD on PMMA.** The selectivity of $TiO_2$ on PMMA determines the potential applications of the AS-ALD process, as each application requires different film thicknesses or number of cycles. Therefore, we study the nucleation delay of $TiO_2$ on PMMA. Figure 1b shows the $TiO_2$ film thickness as a function of the number of ALD cycles. We observe a nucleation delay of 50 cycles on PMMA. The $TiO_2$ film thickness is negligible in the transient regime,[17] and linear growth with a growth per cycle (GPC) of 0.54 Å cycle$^{-1}$ sets in after 50 cycles. In contrast, no nucleation delay occurs on $SiO_2$, where the growth is linear over the whole range with an average GPC of 0.57 Å cycle$^{-1}$. We use the generally accepted definition of the selectivity $S$ from the field of area-selective chemical vapor deposition (Equation 1).[16] Figure 1c shows the accordingly calculated selectivity of the $TiO_2$ ALD process on PMMA. The selectivity is 1 for up to 50 cycles (3.12 nm), decreases rapidly thereafter, and eventually deposition occurs on both surfaces. These results indicate that the difference in nucleation delays on the two surfaces causes AS-ALD of $TiO_2$. The selectivity window is 50 cycles or approximately 3 nm of $TiO_2$ on $SiO_2$. The absolute size of the selectivity window is not readily comparable as it depends, amongst others, on the chemical process reaction, surface chemistry of the substrate, and the deposition parameters. Nevertheless, similar sizes in terms of thicknesses and number of ALD cycles have been observed for dielectric-on-dielectric [5,14,18,19] and metal-on-dielectric [8,11] processes.

The specific surface chemistry of $SiO_2$ and PMMA explains the microscopic nature of the selectivity window in the ALD process as they feature different surface groups with distinct reactivities. $SiO_2$ is primarily –OH terminated and $TiCl_4$ reacts readily in the first self-limiting half reaction

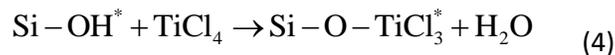

$$\mathrm{Si-OH^{*} + TiCl_4 \rightarrow Si-O-TiCl_3^{*} + H_2O} \quad (4)$$

where the asterisks denote the surface species. In contrast, PMMA features methyl ester as functional group bond to its backbone but no –OH groups. The ALD surface chemistry is more complex with several possible reaction mechanisms: (1) Precursor molecules can infiltrate the polymer and participate in surface-interface reactions, (2) permeate to interstitial spaces and react with the second precursor, or (3) react with other functional groups.[20]

Previous works on the ALD deposition of $Al_2O_3$ and $TiO_2$ on PMMA have found indications for surface-interface reactions,[21] subsurface diffusion,[22,23] reaction with the polymer,[23,24] or full growth inhibition.[6] It is critical to understand the prevalent mechanism to be able to enhance the selectivity of the $TiO_2$ on PMMA during the AS-ALD process. Choosing appropriate, sufficiently-thick polymer film suppress surface-interface reactions by preventing precursors from reaching the interface.[25] Also, prolonging the purge steps removes adsorbed and diffused precursors and decreases subsurface diffusion.[23] Although we use 300 nm thick PMMA films and relatively long purge times, we detected Ti species on the surface of PMMA in X-ray photoelectron spectroscopy (XPS, Figure S3). This indicates that a reaction of TiCl$_4$ with PMMA is the prevalent mechanism for the loss of selectivity. It has been suspected early on that TiCl$_4$ may be complexed by the carbonyl group in PMMA.[25] Indeed, it is well know that hard Lewis acids can easily be coordinated by carbonyl compounds.[26,27] We perform infrared spectroscopy to identify the chemical mechanism for the loss of selectivity on PMMA (Figure 2a). We calculate the differential absorbance to visualize small changes in absorbance (Figure S4). The PMMA film before AS-ALD shows characteristic peaks attributed to the CH$_3$ asymmetric and symmetric stretching at 3000 and 2950 cm$^{-1}$, and the C=O ester and carbonyl stretching features at 1720 and 1260 cm$^{-1}$. After 25 cycles of alternating TiCl$_4$ and H$_2$O pulses, the peak intensity of the C=O ester stretching feature at 1720 cm$^{-1}$ decreases significantly indicating the reaction between TiCl$_4$ and PMMA. The peak intensities of the CH$_3$ stretching features at 3000 and 2950 cm$^{-1}$ decrease likewise but less pronounced. Furthermore, a mode appears near 1400 cm$^{-1}$ that belongs to the lattice vibration of TiO$_2$ (Ti-O-Ti stretching).[28] The reaction of TiCl$_4$ with the PMMA surface can be separated in two different mechanisms. Adsorbed TiCl$_4$ acts catalytically like a Lewis acid activating the ester group to favor the nucleophilic attack of H$_2$O on the C=O carbon yielding HO-CH$_3$ and the carboxylic acid:

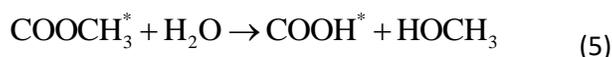

$$\text{COOCH}_3^* + \text{H}_2\text{O} \rightarrow \text{COOH}^* + \text{HOCH}_3 \qquad (5)$$

here, the asterisks denote the surface species. The mass effect of the water pulse is the main driving force of the reaction. Consequently, the growth is zero in this phase as TiCl$_4$ acts as a catalyst (*cf.* selectivity window, Figure 1b). Second, the reaction proceeds with the same mechanism displayed on SiO$_2$ once the hydroxyl on the carboxylic acid has formed:

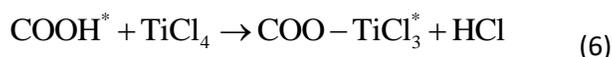

$$\text{COOH}^* + \text{TiCl}_4 \rightarrow \text{COO}-\text{TiCl}_3^* + \text{HCl} \qquad (6)$$

*i.e.,* TiO$_2$ grows on PMMA similar to SiO$_2$ explaining the identical GPC (Figure 1b).

One important feature of PMMA is that the carbonyl group is not located in the backbone of the polymer. Thus, the chain will not dissociate when reacting with TiCl$_4$ making it a robust choice for area deactivation in the AS-ALD processes. The appearance of an additional TiCl$_x$ specie on PMMA in XPS supports this interpretation (Figure S3).

**Surface diffusion in area-selective ALD.** The complex surface chemistry in AS-ALD on polymers yields very different results compared to inorganic substrates. Typically, the topography of the thin film changes,[19] effects like swelling of the polymer[29] or delamination of the thin film occur,[30] and the GPC can change upon patterning.[31] These issues make a precise process control challenging and underline the need for a systematic understanding of the fundamental processes. Therefore, we study the topography and the GPC to facilitate the control of the AS-ALD process.

We measured the topography of PMMA as a function of the number of cycles to examine the formation of $TiO_2$ during the area-selective process (Figure 2c-d). The surface of PMMA is smooth, virtually free of defects and the roughness is low at 360 pm as long as the number of cycles is within the selectivity window (25 cycles). The surface structure starts to change once the number of cycles reaches the edge of the selectivity window (50 cycles). Beyond that, three-dimensional growth sets in and small grains start to appear on the PMMA surface (75 cycles). These eventually start growing laterally and form a complete film (100 cycles). Its roughness increases to 480 pm, which is nearly double that of $TiO_2$ grown on $SiO_2$ (280 pm, cf. Figure S5). The topography of the thickest $TiO_2$ film appears globular with small grains. These results are a consequence of the nucleation mechanism of $TiO_2$ on PMMA. During the first 50 cycles, carboxylic acid forms under the catalytic activity of $TiCl_4$ and no apparent growth of $TiO_2$ on PMMA occurs. Beyond the selectivity window, the carboxylic acid presents a reactive site for nucleation and the small grains indicate the onset of $TiO_2$ crystallization. The formation of small grains and their density correlate with the film crystallinity.[32] Similar surface features occur for $Al_2O_3$ (grown from trimethylaluminum and water) and $TiO_2$ (grown from titanium tetrachloride and water) on other polymer surfaces.[19]

One major potential application of AS-ALD is the self-aligned fabrication of pre-patterned and pre-processed devices. A reasonable target film thickness in these applications is 10 nm.[33] However, typical selectivity windows in AS-ALD are lower.[5,8,11,14,18,19] Growing thin films significantly beyond the selectivity window results in unusable structures, *e.g.,* a comparatively rough film forms on the pre-patterned surface making a lift-off virtually impossible. Therefore, it is interesting to study the influence of the patterning step on the final thin film structure, especially when the film thickness is beyond the selectivity window. We prepared a series of different sized PMMA patterns on $SiO_2$ (Figure S6), deposited 50 and 100 cycles of $TiCl_4/H_2O$ (*i.e.,* within and beyond the selectivity window) on the patterned substrates, followed by the lift-off of PMMA. We examined the resulting $TiO_2$ patterns on $SiO_2$ by AFM, and determined the GPC from the topography. We define the area ratio AR as the ratio of the no-growth to growth area to generalize the discussion of the widths of the growth and no-growth areas (Figure S6).

We find that the GPC strongly increases upon patterning (Figure 2b) and is approximately independent of the area ratio (Figure S7). The GPC on unpatterned substrates is 0.63 Å cycle$^{-1}$ and independent of the number of cycles. Unexpectedly, it depends on the number of cycles on patterned substrates. It is 0.63 Å cycle$^{-1}$ when the number of cycles is 50 (black dots), and increases by 1/3 to 0.83 Å cycle$^{-1}$ when the number of cycles is 100 (red dots). This observation is highly remarkable as the GPC should be independent of the number of cycles in an ALD process operating with saturation characteristics. Commonly, the GPC saturates because of steric hindrance of the ligands or the number of reactive surface sites.[34] In the latter case, when the growth is substrate-inhibited, the GPC is low in the beginning before settling to a constant value. This has been observed in the selective deposition of metallic platinum and ruthenium dots on silicon: the GPC decreases on patterned substrates, which is typically attributed to polymer residue on the substrate.[31] However, this explanation does not hold here as the GPC increases on patterned substrates. Another possible process to be considered is the surface diffusion of species as the chemical natures of the PMMA and SiO$_2$ substrates are different.[35,36]

We analyze the topography of the patterns to gain further insight into the growth mechanism reported here. ALD processes operating with saturation characteristics typically yield large degrees of uniformity. Deviations in the form of gradients may indicate diffusion processes[36,37] and explain the increase in GPC. We etched an 8.9 nm deep trench in SiO$_2$ and deposited 100 cycles of TiCl$_4$/H$_2$O to fill the trench. The AFM image of the resulting structure reveals a flat geometry, *i.e.* a filled trench, with sharp spikes of up to 17 nm running vertically along the interface of the growth and no-growth areas (Figure 3a). Additionally, small grains appear on the surface. The species seem to propagate from the no-growth area towards the growth area and pile up at the interface. To analyze the particle distribution in AS-ALD processes, we employ a kinetic Monte-Carlo model of the surface diffusion of particles. The model assumes a uniform particle distribution in the growth and no-growth areas, and let the particles diffuse until the distribution is in a steady state. The final particle distribution from the kinetic Monte-Carlo model strongly resembles the topography of the patterned samples (Figure 3b). A large number of particles accumulates in the growth area, while only a small number of particles remains in the no-growth area where they form small islands. The accumulation occurs preferably at the interface of the growth to no-growth area, where the distribution exhibits localized spikes in agreement with the experimental findings. The accumulation of particles in the growth area explains the strongly increased GPC on patterned substrates and the pile-up at the edges. Particles diffuse from the no-growth area towards the growth area where they remain near the edges producing sharp spikes. The remaining particles in the no-growth areas act as nucleation centers in the following cycles contributing to a decrease of the selectivity. We further examine the influence of the model parameters on the final particle distribution and find that an increase in the number of iteration steps leads to more particles in the growth area (Figure S8). As the kinetic Monte-Carlo model reproduces the experimental observations, we conclude particle diffusion to be the origin of the increased GPC and pile-up at the interface.

A discussion of the nature of the diffusing particle is essential, as it is not obvious from an ALD process operating with saturation characteristics. In the static model, the first precursor chemisorbs at the surface in the first half reaction, and then reacts with the second precursor in the second half reaction forming the product. All participating species chemisorb at the surface and thus are immobile. In reality, the first precursor may physisorb and diffuse on the substrate during the first half cycle, and the product may diffuse during the second half cycle. However, an exact differentiation of the diffusing specie still remains difficult.[35]

**Control of the surface diffusion and selectivity in area-selective ALD.** One important issue in the application of AS-ALD for the self-aligned fabrication of integrated circuits is the size of the selectivity window. To date, no clear target thickness has been defined,[4] but 10 nm appear reasonable.[33] Further, our results indicate that species bond only weakly to the substrate at the beginning of the ALD process. Consequently, we examine the variation of the precursor purge times as a route to control the surface diffusion of species and improve the process towards larger selectivity. We choose 100 cycles, which is outside of the inherent selectivity window of the ALD process, and vary the purge times from short (10/10 s $TiCl_4$/$H_2O$) to long (60/90 s). Short purge times lead to surface diffusion of weakly bonded species, which results in a pile-up at the growth to no-growth interface (Figure 3c). The AFM image reveals a vertical strap of $TiO_2$ on $SiO_2$. The strap is 1.5 µm wide and exhibits a layered topography, which is evident from the brighter areas. The base of the strap is 8 nm thick and sharp spikes of up to 22 nm appear at the edges (Figure 3d). Uniform planes of larger thickness seem to propagate from these edges. One remarkable feature of those structures is the high aspect ratio of the pile-up. Fabrication of such structures in modern integrated circuits like the fins in FinFET devices requires many processing steps. Exploiting the surface diffusion in area-selective ALD enables the facile fabrication of high aspect ratio nanostructures. Long purge times result in a significantly extended selectivity window (Figure 3d). The $TiO_2$ strap exhibits few defects and no pile-up at the growth to no-growth interface. The film thickness and GPC are unaffected by the purge time, but significantly boosted compared to the unpatterned substrate (Figure 2b). This means that surface diffusion of species occurs and the purge gas removes weakly bonded species on the surface such as physisorbed precursor or product molecules. An extension of the precursor purge times improves the selectivity of the AS-ALD process.

CONCLUSIONS

In conclusion, surface modifications of the substrate strongly manipulate the surface diffusion of species in area-selective deposition. The area-selective atomic layer deposition of $TiO_2$ on poly (methyl methacrylate) (PMMA) and $SiO_2$ yields localized nanostructures with tailored aspect ratios. Controlling the surface diffusion enables such nanostructures as it facilitates the growth rate at the interface of the growth and no-growth areas. This straight-forward strategy may enable new self-aligned fabrication schemes in area-selective deposition. Substrates can be designed to control the surface diffusion and thus deposition.

Further, surface modifications accelerate the growth in the entire growth area by 1/3 as species diffuse from the no-growth to the growth areas. The inherent selectivity of the process of 50 cycles or 3.12 nm is limited by the catalytic activity of $TiCl_4$ during the formation of the carboxylic acid on PMMA. Prolonging the purge times allows the preparation of microstructures with film thicknesses well-beyond the inherent selectivity of the process by removing weakly-adsorbed species and prolonging the diffusion time. This allows to improve the selectivity towards high selectivity processes. Future work should explore the new opportunities that arise from controlling the surface diffusion of species in AS-ALD.

**AUTHOR INFORMATION**

**Corresponding Author**

* E-mail: sangam.chatterjee@physik.uni-giessen.de

**ORCID**

Philip Klement: 0000-0001-7044-713X

Michele Bastianello: 0000-0002-3608-9102

Fabian Michel: 0000-0002-4372-8638

Matthias T. Elm: 0000-0001-7014-5772

Christian Heiliger: 0000-0002-8680-5672

Sangam Chatterjee: 0000-0002-0237-5580

**Author Contributions**

P.K. and S.C. conceived the experiments. P.K. fabricated the samples and performed the measurements, assisted by D.A., M.B. and L.G. D.A. and C.H. developed the KMC model, performed the calculations and analyzed the results. F.M. measured XPS and analyzed the results. P.K., M.B., D.A. and S.C. co-wrote the paper. All authors discussed the results and gave approval to the final version of the manuscript.

**Funding Sources**

This work is funded by the German Research Foundation (DFG) via the collaborative research center SFB 1083. P.K., M.B. and M.T.E. thank the German Federal Ministry of Education and Research (BMBF) for funding of the Nano-MatFutur project NiKo (03XP0093). S.C. acknowledges financial support by the Heisenberg program (CH660/8).

**Notes**

The authors declare no competing financial interest.

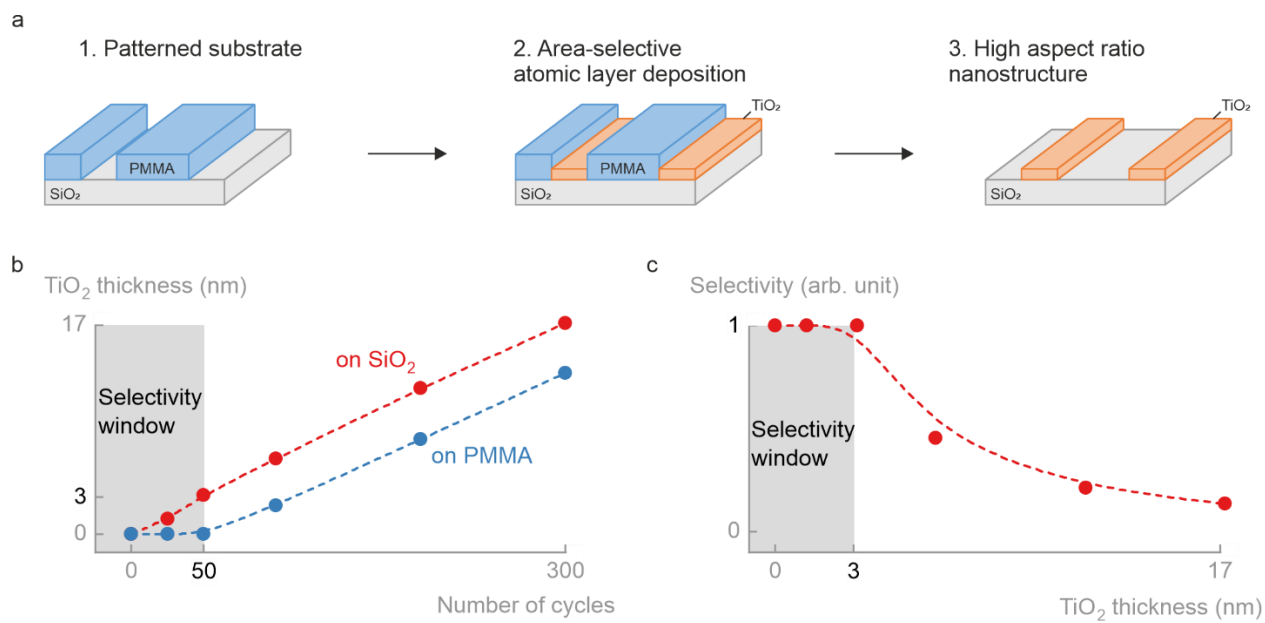

**Figure 1.** Schematic and selectivity of the area-selective ALD approach. a, PMMA patterns on SiO$_2$ serve as starting substrates. TiO$_2$ is deposited in an area-selective ALD process, and mask removal yields high aspect ratio nanostructures. b, TiO$_2$ film thickness on SiO$_2$ and PMMA substrates as a function of the number of cycles. On SiO$_2$, the TiO$_2$ film thickness increases linearly, whereas a nucleation delay of 50 cycles occurs on PMMA. c, Selectivity of the TiO$_2$ deposition on PMMA as a function of the deposited thickness on SiO$_2$. The selectivity window is 3.12 nm or 50 cycles. Dashed lines serve as a guide to the eye.

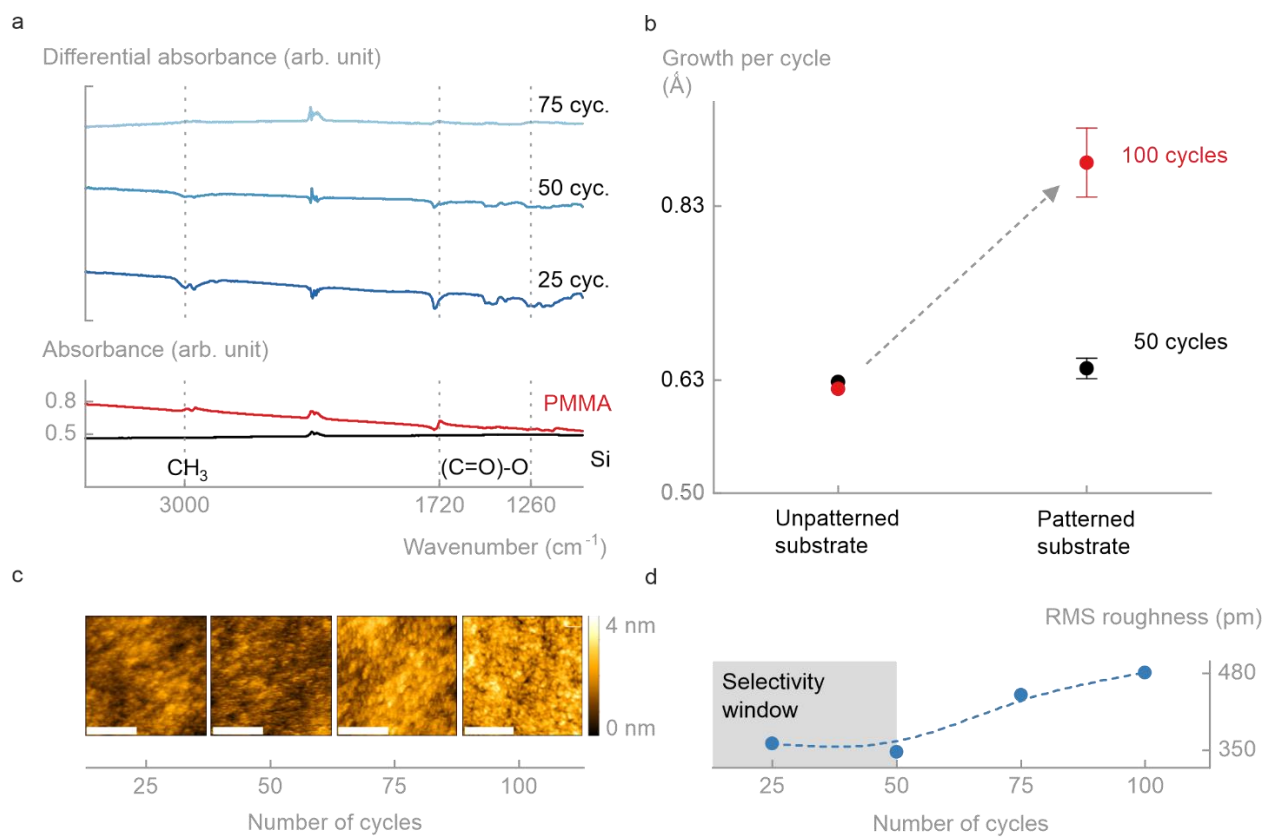

**Figure 2.** FTIR, topography and growth per cycle (GPC) in area-selective ALD. a, Infrared differential absorbance spectra of TiO$_2$ films grown on PMMA. The disappearance of the characteristic CH$_3$ and C=O vibrations near 3000 and 1720 cm$^{-1}$ indicates the nucleophilic attack of H$_2$O on the C=O group of PMMA catalyzed by TiCl$_4$. b, GPC of TiO$_2$ in AS-ALD on (un-)patterned substrates. The GPC on patterned substrates increases significantly when the number of cycles is beyond the selectivity window.
c, Topography of TiO$_2$ on PMMA as a function of the number of cycles. TiO$_2$ islands appear on the surface at 75 cycles and form a complete film at 100 cycles. The scale bar is 200 nm. d, The roughness of TiO$_2$ on PMMA as a function of the number of cycles increases likewise.

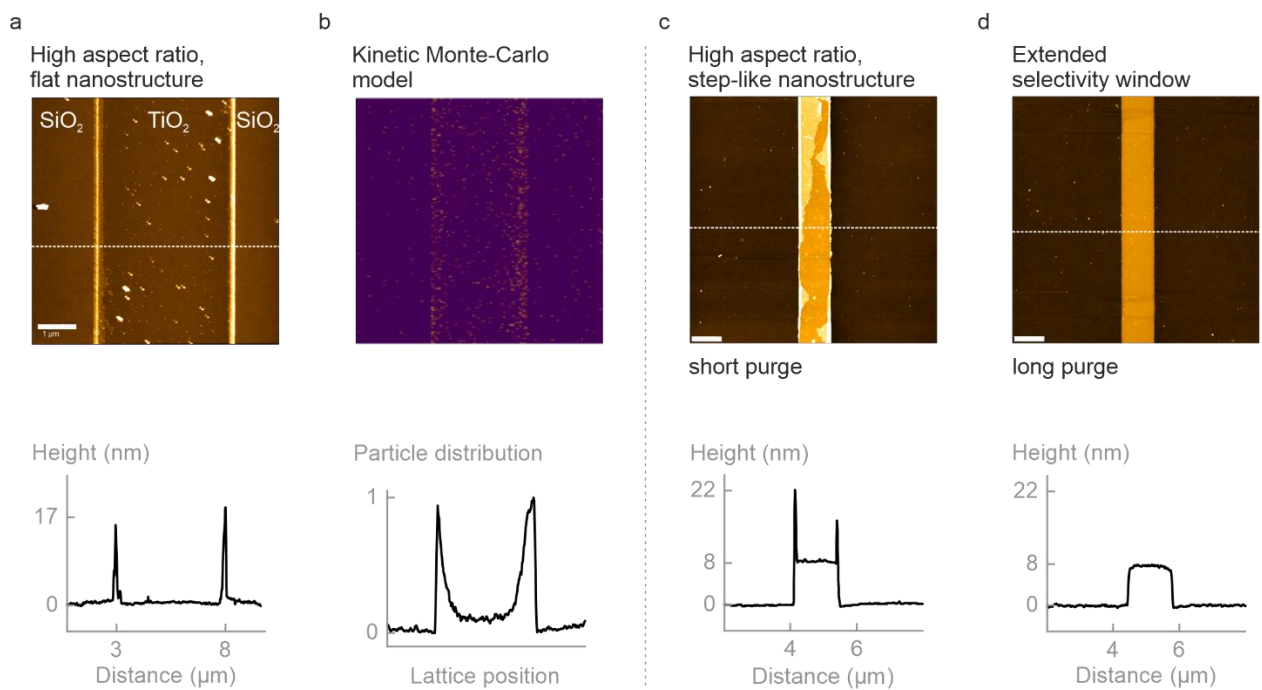

**Figure 3.** Surface diffusion control in area-selective ALD. a, AFM image of a flat TiO$_2$ and SiO$_2$ nanostructure showing pile-up at the edges of the growth and no-growth areas, and corresponding height profile. b, Particle distribution from the kinetic Monte-Carlo model. c, AFM image of a TiO$_2$ strap on SiO$_2$ after 100 cycles with short purge times showing pile-up at the edges and corresponding height profile. d, AFM image of a TiO$_2$ strap on SiO$_2$ after 100 cycles with long purge times and corresponding height profile. The scale bar is 1.5 μm.

# Surface Diffusion Control Enables Tailored Aspect Ratio Nanostructures in Area-Selective Atomic Layer Deposition

Philip Klement, Daniel Anders, Lukas Gümbel, Michele Bastianello, Fabian Michel, Jörg Schörmann, Matthias T. Elm, Christian Heiliger, and Sangam Chatterjee[*]

**Saturation characteristics of the $TiO_2$ ALD process**

We optimized the ALD parameters for the growth of $TiO_2$ on poly (methyl methacrylate) (PMMA) to ensure the saturation characteristics of the process. Therefore, we determined the $TiO_2$ film thickness from X-ray reflectivity measurements. The titanium tetrachloride ($TiCl_4$) pulse time was varied, while the water ($H_2O$) pulse time and the $TiCl_4$ and $H_2O$ purge times were set arbitrarily large to 10 s. This ensures complete saturation of the surface with $H_2O$ and an efficient removal of excess precursors and reaction products from the reactor. The number of cycles was 100, and the growth per cycle (GPC) as a function of the $TiCl_4$ pulse time shows saturation at a pulse duration of 0.1 s (Figure S1A). Then, the $TiCl_4$ pulse time was set to 0.1 s and the $H_2O$ pulse time was varied, while the $TiCl_4$ and $H_2O$ purge times were set arbitrarily large at 10 s. The GPC as a function of the $H_2O$ pulse time shows saturation at a pulse duration of 3 s (Figure S1B).

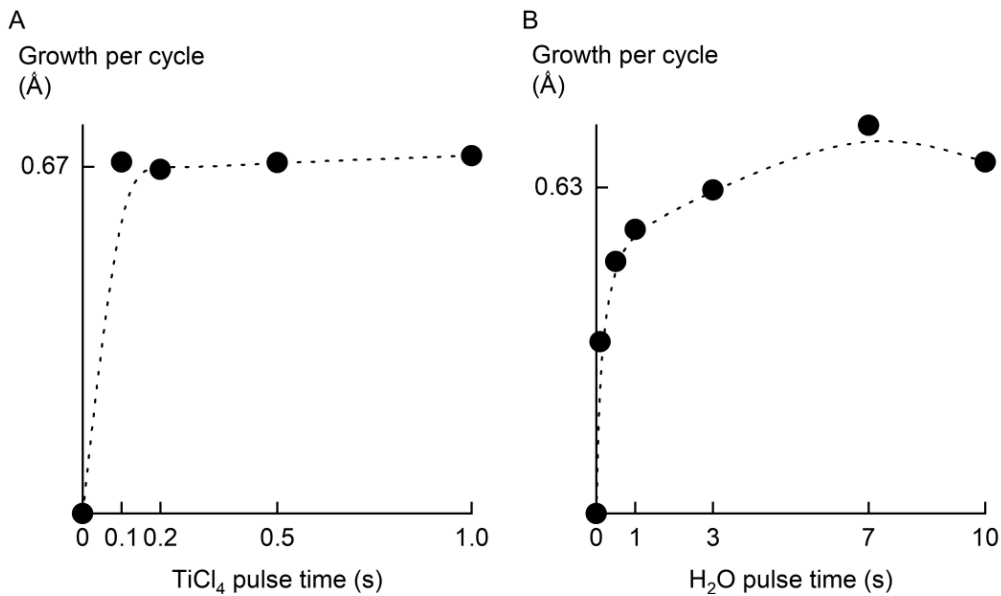

**Figure S1.** Saturation characteristics of the TiO$_2$ ALD process. TiO$_2$ was grown from TiCl$_4$ and H$_2$O on PMMA. (A) GPC of TiO$_2$ as a function of the TiCl$_4$ pulse time. Saturation occurs at 0.1 s. (B) GPC of TiO$_2$ as a function of the H$_2$O pulse time. Saturation occurs at 3 s

**X-ray reflectivity of TiO$_2$ thin films on PMMA**

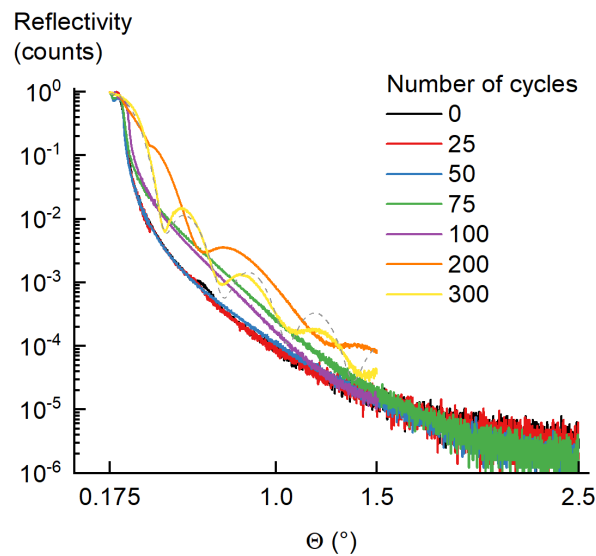

**Figure S2.** X-ray reflectivity of TiO$_2$ thin films on PMMA. TiO$_2$ reflectivity on PMMA (solid lines) as a function of the number of cycles. The dashed grey line is an exemplary simulation of the 300 cycles sample

## X-ray photoelectron spectroscopy

XPS spectra were recorded on PHI VersaProbe with an Al-anode (Al$_{K\alpha}$ = 1486.6 eV) at a source angle of 45°. Charge neutralization of the sample surface was used, and spectra were referenced to the carbon 1s signal (C1s) at 284.8 eV.

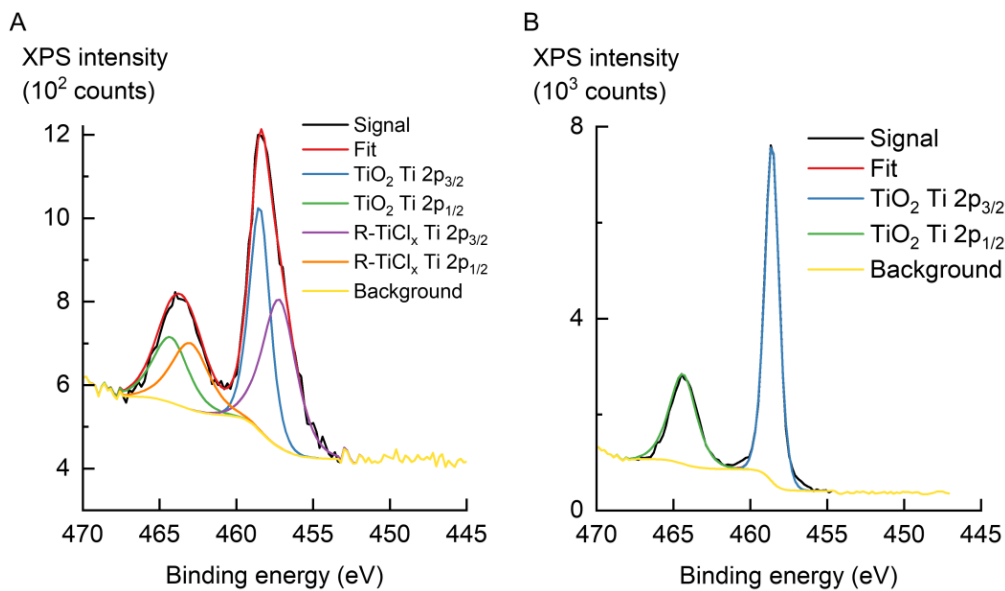

**Figure S3.** X-ray photoelectron spectroscopy of TiO$_2$ thin films. TiO$_2$ signals appear on PMMA (A) and SiO$_2$ (B) indicating growth on both surfaces. The characteristic TiO$_2$ Ti 2p$_{3/2}$ peaks appear at 458.5 eV on SiO$_2$ and PMMA. An additional organic/inorganic TiCl$_2$ compound 2p$_{3/2}$ peak appears on PMMA at a lower energy of 457.3 eV

## Infrared spectroscopy

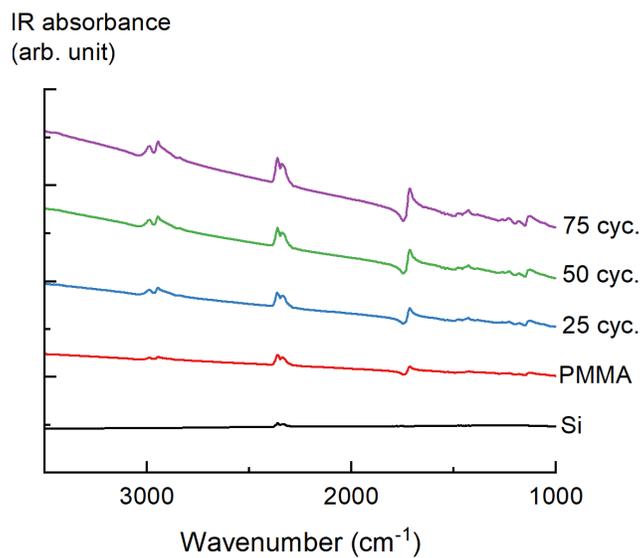

**Figure S4.** Infrared absorbance spectra of $TiO_2$ thin films grown on PMMA. Infrared absorbance of $TiO_2$ thin films as a function of the number of cycles and PMMA and Si references

## Atomic force microscopy

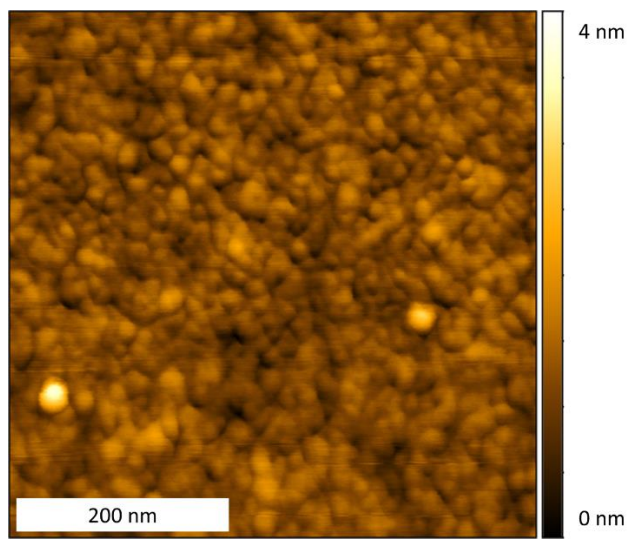

**Figure S5.** Topography of a $TiO_2$ thin film grown on $SiO_2$. The roughness is 280 pm, and the scale bar 200 nm. The number of cycles was 100

**Electron beam lithography patterns**

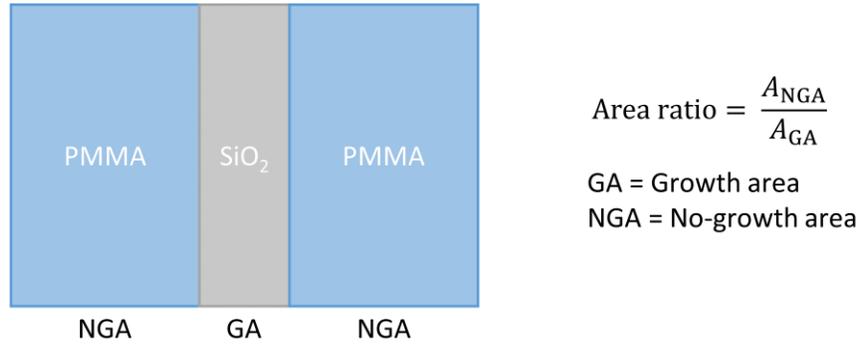

**Figure S6.** Electron beam lithography patterns and area ratio. Schematic representation of the PMMA patterns on SiO$_2$ that serve as substrates in the area-selective ALD process. The ratio of the no-growth area (PMMA) to the growth area (SiO$_2$) was varied and denoted by area ratio

**Growth per cycle and area ratio in AS-ALD**

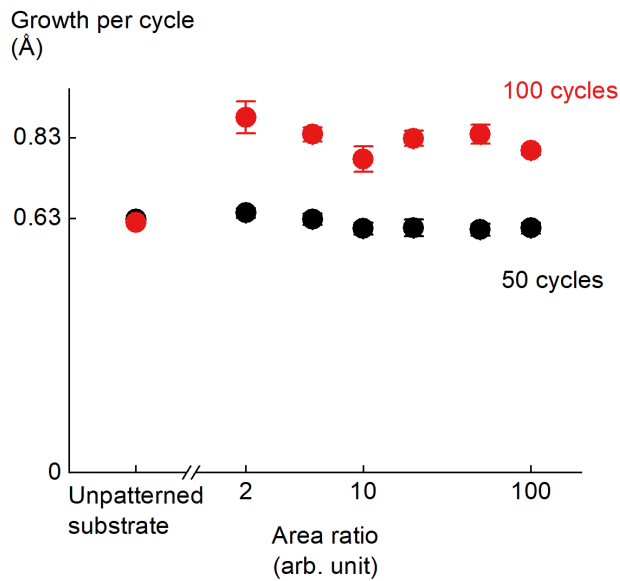

**Figure S7.** Growth per cycle and area ratio in AS-ALD. GPC of TiO$_2$ in AS-ALD on patterned substrates as a function of the area ratio. The GPC on patterned substrates increases significantly when the number of cycles is beyond the selectivity window. Furthermore, it is independent of the area ratio in the examined range

**Iteration steps in kinetic Monte-Carlo**

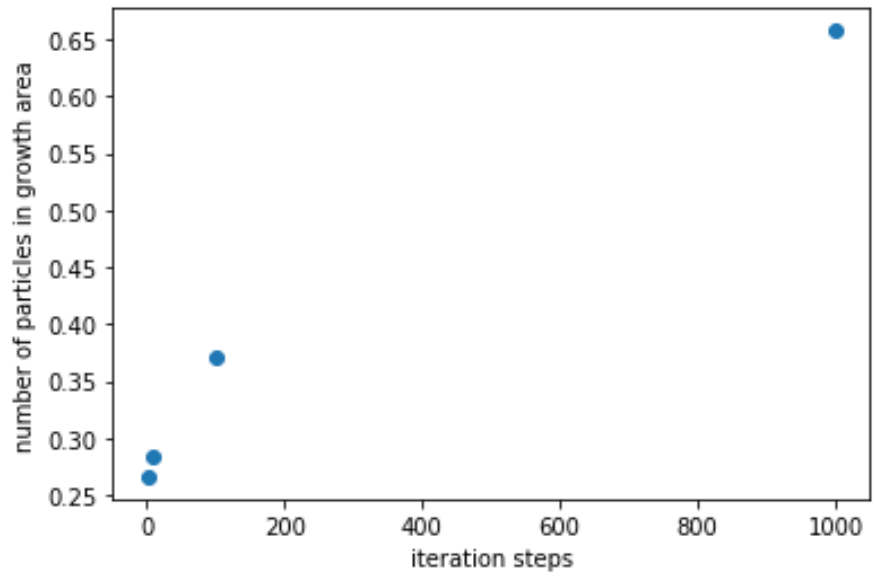

**Figure S8.** Iteration steps in kinetic Monte-Carlo. Number of particles in the growth area as a function of the number of iterations. Particles accumulate in the growth area as the number of iteration steps increases